%% file: top.tex
\documentclass[sigconf]{acmart}

\AtBeginDocument{%
  }

\usepackage{multirow}
\usepackage{booktabs}
\usepackage[ruled,linesnumbered]{algorithm2e}
\usepackage[normalem]{ulem}
\usepackage{balance}
\usepackage{threeparttable}
\usepackage{url}

\usepackage{graphicx}
\usepackage{textcomp}
\usepackage{xcolor}
\usepackage{hyperref}
\usepackage{subcaption}

\newcommand{\methodname}{LevelSyn}

\copyrightyear{2026}
\acmYear{2026}
\setcopyright{cc}
\setcctype{by}
\acmConference[ICCAD '26]{IEEE/ACM International Conference on Computer-Aided Design}{November 08--12, 2026}{San Jose, CA, USA}
\acmBooktitle{IEEE/ACM International Conference on Computer-Aided Design (ICCAD '26), November 08--12, 2026, San Jose, CA, USA}
\acmDOI{10.1145/3831252.3833940}
\acmISBN{979-8-4007-2873-0/2026/11}

\begin{document}

\title{LevelSyn: Physical-Aware Logic Synthesis via Level-Asynchronous Graph Neural Networks}

\author{
Jingyi Zhou, Zhengyuan Shi$^{*}$, Ziyang Zheng, Qiang Xu$^{*}$
}

\affiliation{
\institution{The Chinese University of Hong Kong, Hong Kong, China}
\city{}\country{}
}

\email{{jyzhou26, zyshi21, zyzheng23, qxu}@cse.cuhk.edu.hk}

\renewcommand{\shortauthors}{Jingyi Zhou et al.}

\begin{abstract}
As integrated circuit technology scales into the nanometer regime, the traditional disconnect between logic synthesis and physical design has led to significant PPA (Power, Performance, and Area) degradation and prolonged design closure cycles. Traditional logic synthesis relies on non-physical Wire Load Models (WLMs), while recent spectral-based placement predictors often neglect the inherent hierarchical logic depth and signal flow of netlists, which leads to low-fidelity spatial estimations. To bridge this gap, we propose LevelSyn, a novel physical-aware logic synthesis framework that integrates hierarchical representation learning with a wirelength-driven optimization engine. At its core, LevelSyn leverages a level-asynchronous Graph Neural Network (GNN) to predict high-fidelity gate coordinates by capturing the structural and directional semantics of And-Inverter Graphs (AIGs). To handle industrial-scale designs, a level-aligned subgraph partitioning strategy is introduced to eliminate memory bottlenecks while preserving local logical dependencies. These spatial insights are seamlessly integrated into a newly developed physical-informed synthesis engine within the Berkeley ABC framework. Experimental results on the EPFL benchmark suite demonstrate that LevelSyn significantly outperforms state-of-the-art (SOTA) methods, achieving an average power reduction of 6.89\% and a timing delay improvement of 27.48\%. Furthermore, post-place-and-route validation shows a 99.59\% reduction in design rule check (DRC) violations, highlighting its effectiveness in accelerating design convergence. 

\end{abstract}

\maketitle

\setcounter{footnote}{0}
\begingroup
\renewcommand{\thefootnote}{}
\footnote{Our code is available at \url{https://github.com/zhoujy22/PhysicalAwareSynthesis}.}
\renewcommand{\thefootnote}{*}
\footnotetext{Corresponding authors}
\endgroup

\input{01_introduction_Stone}
\input{02_related}
\input{03_method}
\input{04_experiment}
\input{05_conclusion}

\newpage
\section*{Acknowledgments}
This work was supported in part by the Hong Kong Research Grants Council (RGC) under Grant No. 14202824, 14205925, C6003-24Y, and T46-415/25-R.

\balance
\bibliographystyle{ACM-Reference-Format}
\bibliography{reference}

\end{document}

%% file: 01_introduction_Stone.tex
\section{Introduction}
To avoid local optimization and time-consuming design iterations within the current stage-based chip design workflow, the Electronic Design Automation (EDA) industry has actively adopted a ``shift-left'' strategy~\cite{chen2024large, wu2025shift}. The future design metrics and potential violations should be exposed at much early stages, so that engineers can make revisions on the ``left'' of the workflow. 

Physical-aware synthesis~\cite{pedram1991layout} is a prominent example in logic synthesis domain. Unlike conventional tools~\cite{mishchenko2007abc, wolf2013yosys} estimate the design quality based on models only considering the logical abstraction (e.g., wire load models), Physical-aware synthesis tries to combine the backend information to enable accurate estimation of interconnect delay and congestion during early stages. Although logic synthesis can produce promising results, they are often poorly suited for physical implementation in modern nanometer-scale designs~\cite{lu1998combining, future-of-wires}. To bridge this gap, the early attempts explored the physical-aware synthesis, which incorporates physical information into the early stages of logic synthesis, thereby closing the gap between logical abstraction and physical implementation~\cite{Address_problem, delay_opt_fpga, placement_fpga, dsm_design_flow, Reis2018}. 

Most recently, graph analysis techniques have been employed to predict physical metrics. For example, Net$^2$\cite{xie2021net2} and another attempt~\cite{wu2025differentiable} leverage Graph Neural Networks (GNNs) for pre-placement wirelength estimation. CPA-Remap~\cite{he2025cpa} identifies the critical paths and merges and then remaps the netlist for timing optimization. However, these predictive models are tied to a specific circuit structure. As the logic synthesis constantly reshapes the logic network, these methods are limited to minor refinement after synthesis. 
Other frameworks~\cite{PigMAP1, PigMAP2, he2025cpa} combine the fast physical estimation~\cite{GiFt} into logic optimization and technology mapping tools. Although they can predict the connectivity and spatial distance between logic cells, other critical information, such as hierarchical level structure and directional signal flow, are neglecting. Therefore, efficiently estimating high-fidelity physical information for logic synthesis remains a significant challenge.

In this work, we propose \methodname, a physical-aware logic synthesis framework that leverages a hierarchical neural network to estimate spatial coordinates and orientations, bridging the gap between logical abstraction and physical implementation. Specifically, our framework leverages a single-round and level-asynchronous GNN (LA-GNN) architecture inspired by DeepGate2~\cite{shi2023deepgate2}. We propose novel mechanism to encode and propagate the pre-defined constraints of Primary Inputs (PIs) and Primary Outputs (POs). Unlike existing physical-aware synthesis methods, LevelSyn produces row-consistent spatial priors. In this approach, the cell orientation is directly derived from the predicted $y$-coordinates based on the row specifications in the PDK. This ensures that each cell aligns correctly with the power rails without additional processing. Moreover, to ensure the scalability for industrial-scale designs, we employ a level-aligned subgraph partitioning strategy that mitigates memory overhead. 

We implement our model into an open-source logic synthesis tool ABC~\cite{mishchenko2007abc}. Our model provides the spatial insights of the given pre-mapping logic and guides the technology mapping engine by redefining its cost function. The enhanced ABC tool produces the post-synthesize circuit netlists that are inherently optimized for physical implementation. Using the DREAMPlace~\cite{DreamPlace} for physical design validation, the experimental results on the EPFL benchmark suite demonstrate that LevelSyn outperforms state-of-the-art methods, achieving an average power reduction by 6.89\% and a timing delay reduction by 27.48\%. Furthermore, post-place-and-route validation reveals a 99.59\% reduction in design rule check (DRC) violations, highlighting the framework's effectiveness in generating layout-friendly netlists and accelerating design convergence for advanced chip products.

The contributions of this work is summarized as follows:
\begin{itemize}
    \item We propose a predictive engine based on the Level-Asynchronous GNN (LA-GNN) that explicitly captures the directional signal flow and hierarchical logic depth of And-Inverter Graphs (AIGs). To ensure scalability for industrial-scale designs, we introduce a Level-Aligned Subgraph Partitioning (LASP) strategy that mitigates memory bottlenecks while preserving critical local logical dependencies.
    \item We develop a high-fidelity spatial prediction scheme that estimates gate coordinates and captures the underlying standard-cell row structure. By leveraging these predictions, the framework derives deterministic priors for cell orientation and power rail alignment, providing more comprehensive physical guidance than traditional spectral-based methods.
    \item We provide \methodname, an end-to-end framework combining GNN with open-source logic synthesis tool. By substituting traditional heuristics with a physical-informed cost function, our framework achieves better performance, lower power consumption and lower post-routing DRC violations within the acceptable runtime. 
\end{itemize}

%% file: 02_related.tex
\section{Related Works} 
\begin{figure*}[t]
    \centering
    \includegraphics[width=1\linewidth]{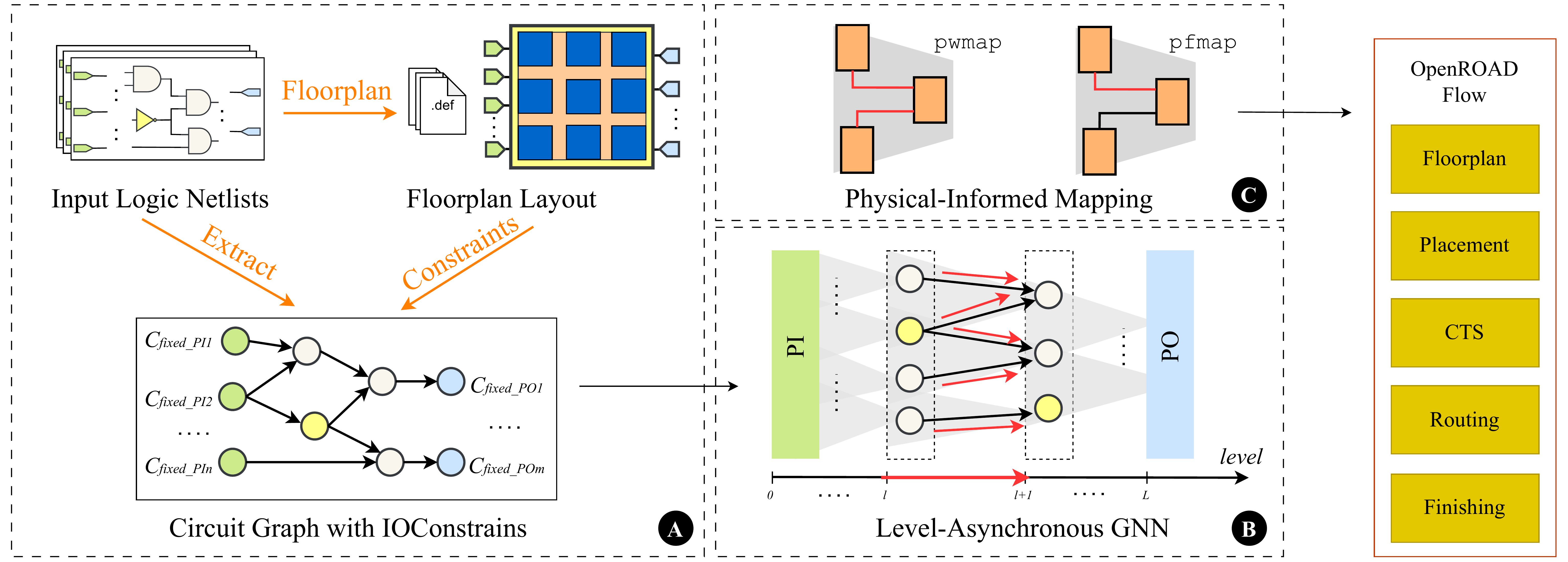}
    \caption{Overview of LevelSyn framework. (A) AIG floorplan to obtain IO constraints. (B) Cell spatial coordinates inference with layer-asynchronous GNN. (C) Physical-informed mapping technique to generate post-mapping netlists.}
    \label{fig:overview}
    \vspace{-10pt}
\end{figure*}
\subsection{Evolution of Physical Aware Synthesis}

Traditionally, VLSI design flows maintained a strict separation between front-end logic synthesis and back-end physical design. However, as semiconductor technology scaled into the deep sub-micron (DSM) era, interconnect delays began to dominate gate delays, rendering traditional logic-depth metrics insufficient for timing closure \cite{future-of-wires}.

To bridge this gap, early research explored the integration of logic optimization and placement. Gosti et al. \cite{Address_problem} addressed timing closure by interleaving logic operations with global placement, while simultaneous technology mapping and placement algorithms like Maple \cite{togawa1994maple, togawa1998maple} sought to incorporate spatial information directly into the mapping process. Despite these efforts, early techniques often suffered from high computational complexity or relied on inaccurate wire-load models (WLM) that failed to capture the nuances of modern routing congestion. Recent industry shifts toward frameworks like OpenROAD \cite{openroad_repo} provide better infrastructure for integration, yet achieving a truly unified metric remains a challenge.

The PigMap series \cite{PigMAP1, PigMAP2} introduced a paradigm shift by utilizing primitive logic gate placement to guide technology mapping. PigMap1 extracts high-fidelity spatial information, specifically estimated coordinates, from a primitive placement of technology-independent gates. This spatial prior allows the mapper to optimize for performance (minimizing critical path Manhattan distance) or power (reducing total wirelength). PigMap2 further refined this by incorporating multi-dimensional metrics such as congestion-aware placement. While successful, these methods rely on fast placement engines that often yields inaccurate spatial data, creating a fidelity gap that prevents the subsequent mapping phase from achieving truly optimal results.

\subsection{GNNs in EDA and Their Limitations}
With the rise of machine learning, Graph Neural Networks (GNNs) have become a popular choice for EDA tasks due to the natural representation of netlists as graphs \cite{chen2024dawn}. Extensive research has focused on learning structural and functional representations for combinational circuits. For instance, DeepGate \cite{li2022deepgate} utilizes attention-based aggregation and skip connections to encode Boolean computations, benefiting downstream tasks such as test point insertion \cite{shi22deeptpi} and Boolean Satisfiability (SAT) \cite{Limin23sat}. This was further evolved by DeepGate2 \cite{shi2023deepgate2}, which introduced pairwise truth-table supervision and functionality-aware loss functions to enhance scalability. Other approaches have explored contrastive learning to derive invariant representations, such as FGNN \cite{deng2024less} for circuit classification, as well as CircuitFusion \cite{fang2025circuitfusion} and NetTAG \cite{fang2025nettag} for multi-modal circuit data. Furthermore, Polargate \cite{liu24polargate} enhanced representation quality by employing the bipolar principle.

Despite these advancements in encoding logical functionality, applying standard GNN architectures to large-scale physical-aware logic synthesis presents significant practical hurdles. First, the primary bottleneck lies in memory consumption. Modern netlists containing millions of gates require massive graph structures and message-passing buffers that frequently exceed the capacity of standard GPU hardware \cite{shi2023deepgate2}. Second, the iterative nature of graph convolutions introduces a substantial runtime overhead compared to traditional heuristics. In a high-speed synthesis environment where rapid design iterations are essential, the inference latency of complex GNNs often outweighs their accuracy benefits \cite{huang2021machine}. Most importantly, while existing representation learning excels at capturing Boolean behavior, it does not natively provide the explicit spatial coordinates required for physical optimization. Therefore, there is a growing need for more efficient neural architectures that can directly provide accurate spatial priors without the prohibitive computational and memory costs associated with standard GNN-based representation learning.

%% file: 03_method.tex
\section{Methodology} \label{Sec:Method}

\subsection{Overview of the Proposed Framework}
The LevelSyn framework is an end-to-end, physical-aware logic synthesis flow designed to bridge the gap between logical abstraction and physical implementation. As illustrated in Fig. ~\ref{fig:overview}, the framework consists of three main phases: Hierarchical Pre-processing, Scalable Spatial Prediction, and Physical-Informed Technology Mapping.

\subsubsection{Problem Formulation}
We represent the input logic netlist as an And-Inverter Graph (AIG), denoted by a directed acyclic graph $G = (V, E)$. Each node $v \in V$ represents a logic gate (2-input AND or Inverter), and each edge $e \in E$ represents the signal flow. Given the fixed boundary constraints from the floorplan stage, specifically the spatial coordinates and the orientations of primary inputs and outputs (PI/POs), denoted as $\mathcal{C}_{fixed} = \{(x_i, y_i, O_i) \mid i \in \text{PI} \cup \text{PO}\}$. Our objective is twofold:
\begin{itemize}
\item  Predict high-fidelity spatial coordinates and capture orientations $(x_v, y_v, O_v)$ for all internal gates $v \in V \setminus (\text{PI} \cup \text{PO})$.
\item Optimize the technology mapping process by utilizing these coordinates to minimize post-routing wirelength and delay.
\end{itemize}

\subsubsection{Framework Pipeline}
The execution flow of LevelSyn follows a structured sequence:

Hierarchical Pre-processing. The framework first parses the AIG to extract topological features and hierarchical level information. To ensure global spatial awareness, fixed PI/PO coordinates are encoded as spatial anchors. For industrial-scale designs, a level-aligned subgraph partitioning strategy is applied to decompose the flattened graph into manageable segments, preventing memory overflow while maintaining logical continuity.

Scalable Spatial Prediction. The partitioned AIG segments are fed into the Level-Asynchronous GNN. Unlike standard GNNs that perform undirected message passing, our engine propagates features along the directional signal flow, which could capture the correlation between logic depth and physical placement. The output is a set of spatial priors $(\hat{x}_v, \hat{y}_v, \hat{O}_v)$ for every gate in the netlist.

Physical-Aware Technology Mapping. These spatial priors are integrated into the Berkeley ABC engine. By extending the mapping cost function to include wirelength-driven metrics, LevelSyn enables the synthesis engine to make physical-aware choices during gate-level optimization, resulting in a netlist with superior PPA.

\begin{figure}[t]
    \centering
    \includegraphics[width=0.9\linewidth]{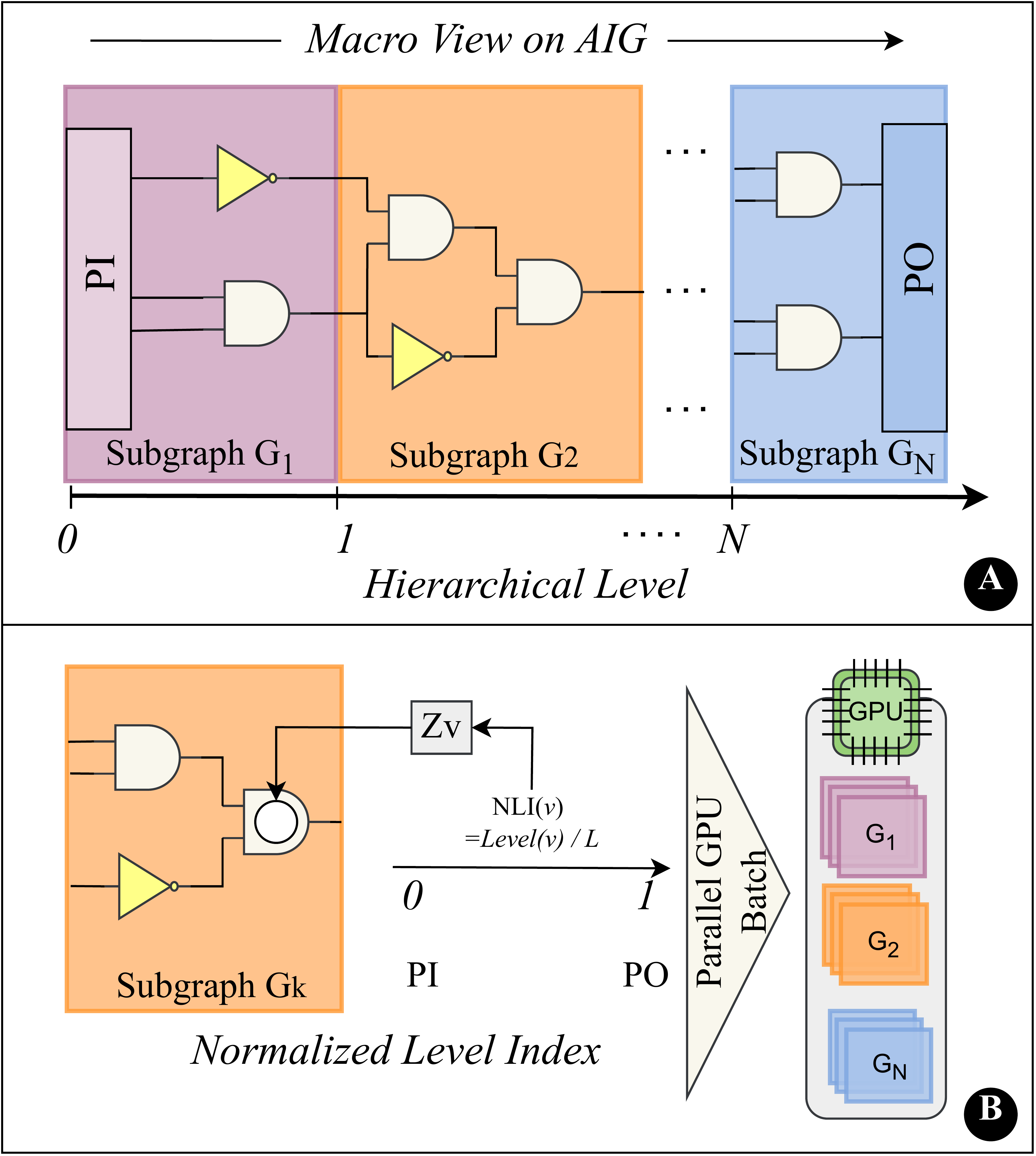}
    \caption{Subgraph partitioning scheme. (A) Level-aligned subgraph partioning overview. (B) Global positional priors injection and parallel GPU batch processing}
    \label{fig:subgraph}
    \vspace{-10pt}
\end{figure}

\subsection{Level-Aware GNN Architecture}
The core of our predictive engine is a Level-Aware Graph Neural Network (LA-GNN) designed to map the logical structure of an AIG to high-fidelity spatial priors. LA-GNN explicitly respects the directional signal flow and hierarchical levels of the netlist.

\subsubsection{Level-Asynchronous Message Passing}
To capture the logic depth and signal propagation, we first partition the AIG nodes into sequential levels $\{L_1, L_2, \dots, L_D\}$, where $D$ is the maximum logic depth. A node $v$ belongs to level $L_k$ if the maximum path length from any PI to $v$ is $k$.

The initial feature vector $z_v$ for each node $v \in V$ is constructed by concatenating structural, functional, and spatial attributes to provide a rich starting point for the level-asynchronous message passing. Specifically, $z_v$ is defined as:
\begin{equation}
z_v = [T_v \parallel S_v \parallel D_v]
\end{equation}
where $\parallel$ denotes the concatenation operator. The components are defined as follows:

\textbf{Gate Type Encoding} ($T_v$): We use a one-hot encoding vector to represent the functional type of the node. 

\textbf{Spatial Anchors} ($S_v$): This is a critical feature for shift-left physical awareness. For nodes $v \in \text{PI} \cup \text{PO}$, $S_v$ contains their fixed $(x, y)$ coordinates obtained from the floorplan stage, normalized to the range $[0, 1]$. For internal logic gates whose locations are unknown, $S_v$ is initialized with the center of the layout $(0.5, 0.5)$ for a place-holder value. This allows the GNN to treat PI/POs as spatial anchors that guide the coordinate regression of internal nodes. 

\textbf{Structural Attributes} ($D_v$): To capture the local complexity of the netlist, we include the fan-in and fan-out degrees of each node. Additionally, the pre-calculated topological level index is embedded as a numerical feature, helping the GNN perceive its relative depth within the entire signal path.

To ensure that every internal gate perceives the spatial constraints from both PIs and POs, we propose a Bi-directional Level-Asynchronous Message Passing mechanism. This approach consists of a forward pass to capture signal flow and a backward pass to diffuse spatial constraints from the outputs.

\textbf{Forward Signal Propagation.} We first propagate features along the topological order from level $L_1$ to $L_D$. For a node $v \in L_k$, the forward hidden state $\overrightarrow{h}_v^{(k)}$ captures the logical influence from its fan-in:

\begin{equation}
\overrightarrow{m}_v^{(k)} = \text{AGGREGATE} \left( { \overrightarrow{h}_u^{(j)} \mid u \in \text{FI}(v), j < k } \right)
\end{equation}

\begin{equation}
\overrightarrow{h}v^{(k)} = \sigma \left( \mathbf{W}{fwd} \cdot \text{CONCAT}(\overrightarrow{m}_v^{(k)}, z_v) \right)
\end{equation}

\textbf{Backward Constraint Diffusion.} To incorporate PO spatial constraints, a second pass is performed in reverse topological order ($L_D$ to $L_1$). For a node $v \in L_k$, its backward hidden state $\overleftarrow{h}_v^{(k)}$ is updated by its fan-out nodes (FO):
\begin{equation}
\overleftarrow{m}_v^{(k)} = \text{AGGREGATE} \left( { \overleftarrow{h}_u^{(j)} \mid u \in \text{FO}(v), j > k } \right)
\end{equation}
\begin{equation}
\overleftarrow{h}v^{(k)} = \sigma \left( \mathbf{W}{bwd} \cdot \text{CONCAT}(\overleftarrow{m}_v^{(k)}, z_v) \right)
\end{equation}

\textbf{Final Embedding Fusion.} The final high-dimensional embedding $h_v^*$ is the concatenation of both directional states: 
\begin{equation}
    h_v^* = [\overrightarrow{h}_v^{(k)} \parallel \overleftarrow{h}_v^{(k)}]
\end{equation}. This dual-flow architecture ensures that the structural representation of each gate is constrained by PI and PO anchors, allowing the downstream MLP to interpolate coordinates that respect the global floorplan.

\subsubsection{Orientation Estimation}
In a standard-cell-based layout, a cell's orientation is not an independent variable but is often constrained by its vertical position relative to the power rails. We derive the orientation estimation $\hat{O}_v$ based on the predicted $\hat{y}_v$ and the row structure defined in the floorplan. First, we map the predicted $\hat{y}_v$ to a discrete row index $R_v$:
\begin{equation}
R_v = \text{round} \left( \frac{\hat{y}_v - y_{offset}}{H_{row}} \right)
\end{equation}
where $H_{row}$ is the standard cell row height and $y_{offset}$ is the layout boundary. Given the power rail sharing property (VDD-VSS-VSS-VDD), the orientation $\hat{O}_v$ is estimated as:
\begin{equation}
\hat{O}_v = \begin{cases} \text{N}, & \text{if } R_v \mod 2 = 0 \\ \text{FS}, & \text{if } R_v \mod 2 = 1 \end{cases}
\end{equation}
This mechanism ensures that the predicted spatial priors $(\hat{x}_v, \hat{y}_v, \hat{O}_v)$ are physically consistent with the target layout environment, allowing the synthesis engine to better estimate the pin locations and parasitic effects during the mapping phase.

\subsection{Scalable Subgraph Partitioning}
To ensure the scalability of LevelSyn on modern designs while preserving the directional logic dependencies, we propose a Level-Aligned Subgraph Partitioning (LASP) strategy (Figure\ref{fig:subgraph}).

Unlike traditional graph partitioning algorithms \cite{graph-partition, Cluster-GNN} that focus on minimizing edge cuts but often disrupt the topological flow, our LASP strategy respects the hierarchical levels of the AIG. Given an AIG with a maximum logic depth $L$, we divide the graph into $N$ contiguous segments along the level dimension. Each subgraph $G_k = (V_k, E_k)$ is defined as:
\begin{equation}
V_k = \{ v \in V \mid (k-1) \cdot w \le \text{Level}(v) < k \cdot w \}    
\end{equation}
where $w = \lceil L/N \rceil$ is the level window size. This partitioning ensures that each subgraph contains a complete "logic slice" of the circuit, maintaining the local connectivity between adjacent logic levels which is critical for spatial proximity estimation.

A key challenge in independent subgraph partitioning is the loss of global coordinates. Instead of using hardware-inefficient sequential propagation, LevelSyn injects Global Positional Priors directly into the initial node features $z_v$. We define a Normalized Level Index (NLI) for each node:
\begin{equation}
\text{NLI}(v) = \frac{\text{Level}(v)}{L}    
\end{equation}
The NLI serves as a global GPS for the GNN, allowing each subgraph—processed in parallel within a GPU batch to perceive its relative distance between the PI (NLI $\approx 0$) and PO (NLI $\approx 1$) boundaries. By combining NLI with the fixed PI/PO spatial anchors, the model can maintain a globally consistent coordinate system.

By controlling the window size $w$, the memory complexity of each GNN pass is reduced from $O(|V|)$ to $O(|V|/N)$, effectively capping the peak memory usage regardless of the total circuit size. This enables LevelSyn to handle industrial benchmarks that are orders of magnitude larger than those typically supported by spectral-based methods, which often suffer from the cubic complexity of Laplacian eigen-decomposition.

\begin{figure}[t]
    \centering
    \includegraphics[width=0.9\linewidth]{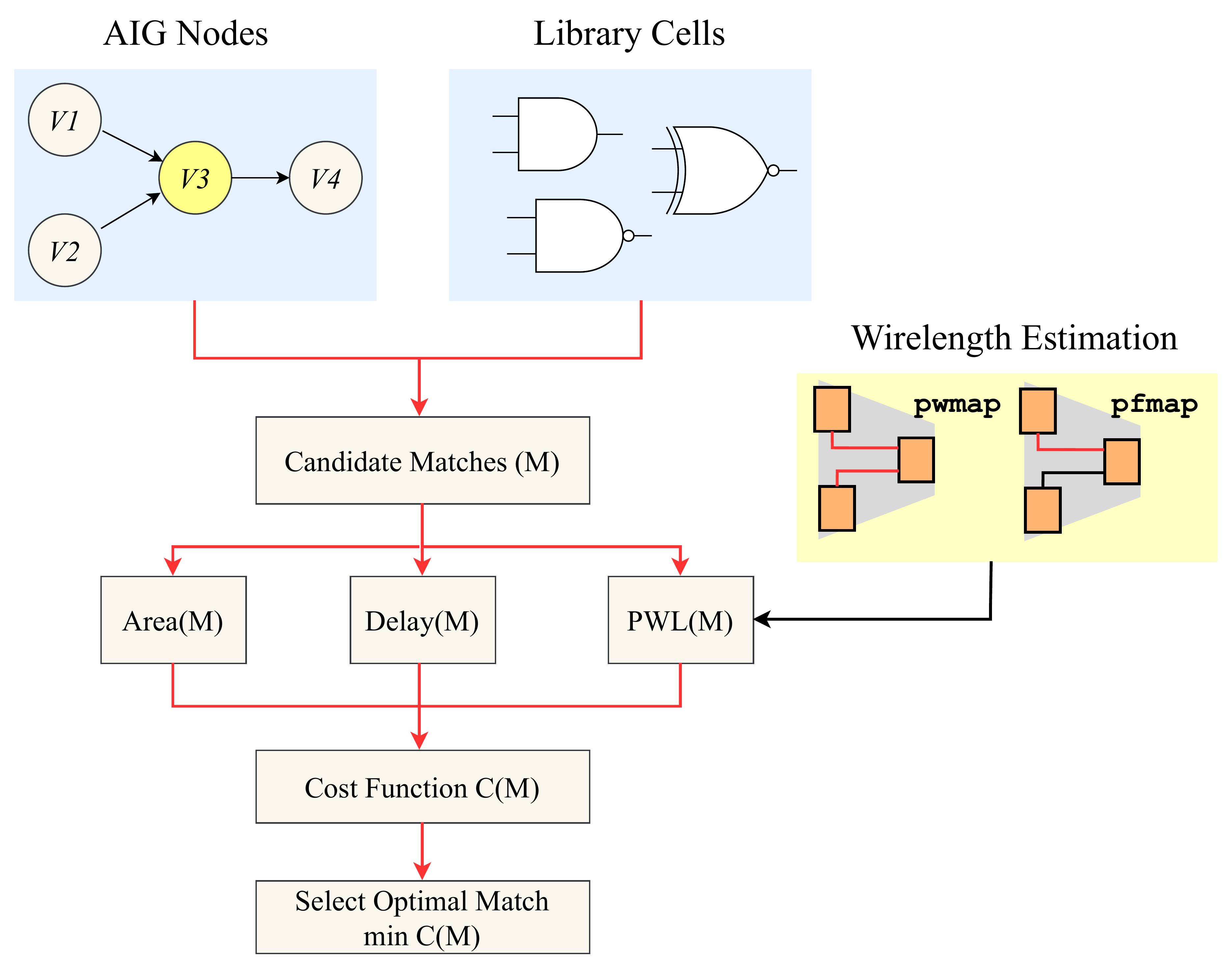}
    \caption{Explanation of the technology mapping flow}
    \label{fig:physical}
    \vspace{-10pt}
\end{figure}

\subsection{Physical-Aware Technology Mapping}
Figure ~\ref{fig:physical} illustrates the final stage of the LevelSyn framework. It integrates the predicted spatial priors into the technology mapping process within the Berkeley ABC \cite{brayton2010abc} engine. Drawing inspiration from the physical-aware synthesis paradigm established by PigMAP \cite{PigMAP1}, we bridge the gap between logical structural optimization and physical layout requirements.

\subsubsection{Physical Data Integration}
To enable physical awareness, the predicted coordinates $(\hat{x}_v, \hat{y}_v)$ and orientations $\hat{O}_v$ for every AIG node are back-annotated into the ABC data structure. We extend the AIG node representation in ABC to include a physical attribute vector $\mathcal{P}_v = (\hat{x}_v, \hat{y}_v, \hat{O}_v)$. During the mapping phase, these attributes are used to estimate the wirelength of potential subject graphs (matches) before they are committed to the final netlist.

\subsubsection{Wirelength-Driven Cost Function}
In standard technology mapping, the engine selects a library gate (cell) to cover a portion of the AIG by minimizing a cost function typically focused on area or arrival time. LevelSyn redefines this cost function to include a Predicted Wirelength (PWL) metric. Following the differentiated cost evaluation strategy proposed in PigMAP \cite{PigMAP1}, we adapt the PWL calculation based on the specific optimization target to better reflect the underlying physical impact. For a candidate match $M$ that implements a set of AIG nodes, the augmented cost $\mathcal{C}(M)$ is defined as:
\begin{equation}
\mathcal{C}(M) = \alpha \cdot \text{Area}(M) + \beta \cdot \text{Delay}(M) + \gamma \cdot \text{PWL}(M)
\end{equation}
where $\alpha, \beta, \gamma$ are weighting factors. The $\text{PWL}(M)$ is estimated using the Manhattan distance between the predicted center of the candidate cell $v_M$ and its fan-in/fan-out neighbors. 
For power-oriented command \textbf{pwmap}, we calculated the sum of Manhattan distance to all adjacent pins to minimize total switching power:
\begin{equation}
\text{PWL}(M) = \sum_{u \in \text{adj}(v_M)} (|\hat{x}_{v_M} - \hat{x}_u| + |\hat{y}_{v_M} - \hat{y}_u|)
\end{equation}
Conversely, for delay-oriented mapping command \textbf{pfmap}, we employ the maximum Manhattan distance to bound the worst-case interconnect delay on critical paths:
\begin{equation}
\text{PWL}(M) = \max_{u \in \text{adj}(v_M)} (|\hat{x}_{v_M} - \hat{x}_u| + |\hat{y}_{v_M} - \hat{y}_u|)
\end{equation}

\begin{table*}[!t]
\centering
\caption{PPA Comparison under Different Technology Mapping Strategies}
\label{tab:main_result}
\begin{threeparttable}
\resizebox{\linewidth}{!}{
\begin{tabular}{l c  ccc  ccc  ccc}
\toprule
\multirow{2}{*}{\textbf{Circuit}} & \multirow{2}{*}{\textbf{Nodes}} & \multicolumn{3}{c}{\textbf{Baseline}} & \multicolumn{3}{c}{\textbf{\textit{PigMap-power}}} & \multicolumn{3}{c}{\textbf{\textit{PigMap-performance}}} \\
\cmidrule(lr){3-5} \cmidrule(lr){6-8} \cmidrule(lr){9-11}
& & \textbf{Area($um^2$)} & \textbf{Delay($ps$)} & \textbf{Power($mW$)} & \textbf{Area($um^2$)} & \textbf{Delay($ps$)} & \textbf{Power($mW$)} & \textbf{Area($um^2$)} & \textbf{Delay($ps$)} & \textbf{Power($mW$)} \\
\midrule
ctrl & 181 & 571.80 & 760.23 & 0.0063 & 604.33 & 737.59 & 0.0089 & 539.27 & 826.13 & 0.0078 \\
int2float & 271 & 938.40 & 886.41 & 0.0084 & 1,068.52 & 1,068.23 & 0.0015 & 829.55 & 1,045.27 & 0.0110 \\
dec & 312 & 1,216.17 & 712.31 & 0.0226 & 1,376.32 & 1,257.90 & 0.0292 & 1,376.32 & 1,257.90 & 0.0292 \\
router & 317 & 1,657.84 & 3,685.60 & 0.0236 & 1,356.30 & 4,086.01 & 0.0201 & 1,177.38 & 3,214.47 & 0.0167 \\
cavlc & 703 & 2,858.99 & 1,391.17 & 0.0306 & 2,870.25 & 1,718.93 & 0.0467 & 2,312.22 & 1,493.83 & 0.0353 \\
priority & 1,106 & 6,586.32 & 17,277.00 & 0.0993 & 4,713.27 & 14,011.38 & 0.0693 & 5,268.80 & 17,847.96 & 0.0773 \\
adder & 1,276 & 7,573.51 & 19,252.33 & 0.0933 & 3,921.26 & 17,057.93 & 0.0511 & 4,128.96 & 18,386.91 & 0.0552 \\
i2c & 1,489 & 5,615.39 & 1,365.56 & 0.0623 & 5,874.38 & 1,838.38 & 0.0847 & 4,919.72 & 1,405.88 & 0.0681 \\
max & 3,377 & 18,470.21 & 41,957.09 & 0.2390 & 13,890.82 & 30,964.82 & 0.2150 & 13,346.55 & 28,368.59 & 0.2100 \\
bar & 3,471 & 10,959.26 & 4,993.18 & 0.1750 & 18,342.59 & 8,062.43 & 0.3260 & 15,162.04 & 7,235.70 & 0.2490 \\
sin & 5,440 & 43,845.80 & 27,658.14 & 0.9000 & 27,699.06 & 20,565.40 & 0.5240 & 29,926.20 & 30,132.74 & 0.5800 \\
arbiter & 12,095 & 40,650.23 & 8,641.04 & 0.6710 & 46,036.65 & 8,170.64 & 0.8970 & 38,745.91 & 7,782.34 & 0.7440 \\
voter & 14,759 & 139,540.08 & 8,000.55 & 2.3200  & 91,849.34 & 7,312.21 & 1.6200  & 93,568.48 & 7,970.67 & 1.6900  \\
square & 18,548 & 148,728.89 & 20,094.88 & 2.4500  & 88,584.96 & 17,836.24 & 1.4700  & 91,456.46 & 18,908.50 & 1.4800  \\
sqrt & 24,746 & 204,521.14 & 1,517,812.00 & 4.5300  & 143,030.92 & 789,714.62 & 2.8900  & 130,052.23 & 861,341.06 & 2.6000  \\
multiplier & 27,190 & 204,043.19 & 29,332.91 & 3.9100  & 133,090.14 & 24,757.79 & 2.3800  & 133,267.81 & 28,860.17 & 2.3900  \\
log2 & 32,092 & 260,310.91 & 168,470.03 & 5.2700  & 150,314.16 & 42,295.55 & 2.7700  & 149,226.86 & 49,459.25 & 2.7900  \\
mem\_ctrl & 48,040 & 248,791.11 & 68,602.83 & 4.0700  & 203,206.14 & 24,483.87 & 3.5000  & 173,094.75 & 19,936.07 & 2.9800  \\
div & 57,375 & 523,522.09 & 690,338.06 & 10.7000 & 280,609.12 & 399,128.59 & 5.1100  & 277,827.69 & 436,291.28 & 5.1700  \\
hyp & 214,591 & 1,760,069.25 & 5,888,406.50 & 33.2000 & 1,171,167.00 & 4,321,252.00 & 20.3000 & 1,203,100.12 & 4,261,045.00 & 20.1000 \\
\midrule
\midrule
\multirow{2}{*}{\textbf{Circuit}} & \multirow{2}{*}{\textbf{Nodes}} & & \multicolumn{3}{c}{\textbf{\textit{LevelSyn-pwmap} (Ours)}} & \multicolumn{3}{c}{\textbf{\textit{LevelSyn-pfmap} (Ours)}} & \multicolumn{2}{c}{} \\
\cmidrule(lr){4-6} \cmidrule(lr){7-9}
& & & \textbf{Area ($Imp.^\dagger$)} & \textbf{Delay ($Imp.^\dagger$)} & \textbf{Power ($Imp.^\dagger$)} & \textbf{Area ($Imp.^\ddagger$)} & \textbf{Delay ($Imp.^\ddagger$)} & \textbf{Power ($Imp.^\ddagger$)} & \multicolumn{2}{c}{} \\
\midrule
ctrl & 181& & 601.83 {\scriptsize (0.4\%)} & 741.25 {\scriptsize (-0.5\%)} & 0.0088 {\scriptsize (1.5\%)} & 519.25 {\scriptsize (3.7\%)} & 799.10 {\scriptsize (3.3\%)} & 0.0076 {\scriptsize (2.3\%)} & \multicolumn{2}{c}{} \\
int2float  & 271 & & 1,042.25 {\scriptsize (2.5\%)} & 1,113.13 {\scriptsize (-4.2\%)} & 0.0151 {\scriptsize (0.7\%)} & 825.79 {\scriptsize (0.5\%)} & 1,009.60 {\scriptsize (3.4\%)} & 0.0109 {\scriptsize (0.9\%)} & \multicolumn{2}{c}{} \\
dec & 312& & 1,376.32 {\scriptsize (0.0\%)} & 1,257.90 {\scriptsize (0.0\%)} & 0.0283 {\scriptsize (3.1\%)} & 1,396.34 {\scriptsize (-1.5\%)} & 1,257.90 {\scriptsize (0.0\%)} & 0.0295 {\scriptsize (-1.0\%)} & \multicolumn{2}{c}{} \\
router & 317 & & 1,307.50 {\scriptsize (3.6\%)} & 4,162.74 {\scriptsize (-1.9\%)} & 0.0194 {\scriptsize (3.5\%)} & 1,174.88 {\scriptsize (0.2\%)} & 3,213.78 {\scriptsize (0.0\%)} & 0.0167 {\scriptsize (0.0\%)} & \multicolumn{2}{c}{} \\
cavlc & 703 && 2,818.95 {\scriptsize (1.8\%)} & 1,510.57 {\scriptsize (12.1\%)} & 0.0439 {\scriptsize (6.0\%)} & 2,324.73 {\scriptsize (-0.5\%)} & 1,516.30 {\scriptsize (-1.5\%)} & 0.0354 {\scriptsize (-0.3\%)} & \multicolumn{2}{c}{} \\
priority & 1,106 && 4,839.64 {\scriptsize (-2.7\%)} & 13,963.57 {\scriptsize (0.3\%)} & 0.0710 {\scriptsize (-2.4\%)} & 5,637.91 {\scriptsize (-7.0\%)} & 18,074.93 {\scriptsize (-1.3\%)} & 0.0826 {\scriptsize (-6.9\%)} & \multicolumn{2}{c}{} \\
adder & 1,276 && 3,856.20 {\scriptsize (1.7\%)} & 17,020.18 {\scriptsize (0.2\%)} & 0.0493 {\scriptsize (3.5\%)} & 4,331.65 {\scriptsize (-4.9\%)} & 16,410.26 {\scriptsize (10.8\%)} & 0.0609 {\scriptsize (-10.3\%)} & \multicolumn{2}{c}{} \\
i2c & 1,489 && 5,873.13 {\scriptsize (0.0\%)} & 2,293.93 {\scriptsize (-24.8\%)} & 0.0855 {\scriptsize (-0.9\%)} & 4,918.47 {\scriptsize (0.0\%)} & 1,541.01 {\scriptsize (-9.6\%)} & 0.0681 {\scriptsize (0.0\%)} & \multicolumn{2}{c}{} \\
max & 3,377 && 16,429.51 {\scriptsize (-18.3\%)} & 22,742.97 {\scriptsize (26.6\%)} & 0.2610 {\scriptsize (-21.4\%)} & 13,663.10 {\scriptsize (-2.4\%)} & 20,788.38 {\scriptsize (26.7\%)} & 0.2110 {\scriptsize (-0.5\%)} & \multicolumn{2}{c}{} \\
bar & 3,471 && 13,387.84 {\scriptsize (27.0\%)} & 5,719.02 {\scriptsize (29.1\%)} & 0.2460 {\scriptsize (24.5\%)} & 12,092.85 {\scriptsize (20.2\%)} & 6,473.61 {\scriptsize (10.5\%)} & 0.2000 {\scriptsize (19.7\%)} & \multicolumn{2}{c}{} \\
sin & 5,440 && 29,171.73 {\scriptsize (-5.3\%)} & 19,504.01 {\scriptsize (5.2\%)} & 0.5550 {\scriptsize (-5.9\%)} & 29,428.22 {\scriptsize (1.7\%)} & 22,205.18 {\scriptsize (26.3\%)} & 0.5770 {\scriptsize (0.5\%)} & \multicolumn{2}{c}{} \\
arbiter & 12,095 && 45,880.25 {\scriptsize (0.3\%)} & 8,447.16 {\scriptsize (-3.4\%)} & 0.8960 {\scriptsize (0.1\%)} & 40,267.37 {\scriptsize (-3.9\%)} & 6,493.06 {\scriptsize (16.6\%)} & 0.6810 {\scriptsize (8.5\%)} & \multicolumn{2}{c}{} \\
voter & 14,759& & 95,342.69 {\scriptsize (-3.8\%)} & 7,585.59 {\scriptsize (-3.7\%)} & 1.6900  {\scriptsize (-4.3\%)} & 95,544.13 {\scriptsize (-2.1\%)} & 7,429.42 {\scriptsize (6.8\%)} & 1.7300  {\scriptsize (-2.4\%)} & \multicolumn{2}{c}{} \\
square & 18,548& & 85,638.38 {\scriptsize (3.3\%)} & 17,756.67 {\scriptsize (0.4\%)} & 1.4400  {\scriptsize (2.0\%)} & 83,546.38 {\scriptsize (8.6\%)} & 18,270.79 {\scriptsize (3.4\%)} & 1.3700  {\scriptsize (7.4\%)} & \multicolumn{2}{c}{} \\
sqrt & 24,746 && 156,964.28 {\scriptsize (-9.7\%)} & 739,768.12 {\scriptsize (6.3\%)} & 3.0600  {\scriptsize (-5.9\%)} & 158,259.28 {\scriptsize (-21.7\%)} & 752,540.00 {\scriptsize (12.6\%)} & 3.1700  {\scriptsize (-21.9\%)} & \multicolumn{2}{c}{} \\
multiplier & 27,190 && 134,481.47 {\scriptsize (-1.1\%)} & 38,750.29 {\scriptsize (-56.5\%)} & 2.3700  {\scriptsize (0.4\%)} & 141,235.45 {\scriptsize (-6.0\%)} & 37,628.16 {\scriptsize (-30.4\%)} & 2.5900  {\scriptsize (-8.4\%)} & \multicolumn{2}{c}{} \\
log2 & 32,092& & 155,928.30 {\scriptsize (-3.7\%)} & 48,510.23 {\scriptsize (-14.7\%)} & 2.8900  {\scriptsize (-4.3\%)} & 162,428.28 {\scriptsize (-8.8\%)} & 50,786.61 {\scriptsize (-2.7\%)} & 3.0500  {\scriptsize (-9.3\%)} & \multicolumn{2}{c}{} \\
mem\_ctrl & 48,040& & 204,930.28 {\scriptsize (-0.8\%)} & 19,050.44 {\scriptsize (22.2\%)} & 3.4500  {\scriptsize (1.4\%)} & 175,667.22 {\scriptsize (-1.5\%)} & 18,280.53 {\scriptsize (8.3\%)} & 2.9400  {\scriptsize (1.3\%)} & \multicolumn{2}{c}{} \\
div & 57,375 && 274,719.72 {\scriptsize (2.1\%)} & 428,615.28 {\scriptsize (-7.4\%)} & 5.0200  {\scriptsize (1.8\%)} & 281,354.84 {\scriptsize (-1.3\%)} & 464,370.06 {\scriptsize (-6.4\%)} & 5.2000  {\scriptsize (-0.6\%)} & \multicolumn{2}{c}{} \\
hyp & 214,591 && 1,094,646.12 {\scriptsize (6.5\%)} & 3,133,495.00 {\scriptsize (27.5\%)} & 18.9000 {\scriptsize (6.9\%)} & 1,238,242.50 {\scriptsize (-2.9\%)} & 3,304,790.25 {\scriptsize (22.4\%)} & 21.8000 {\scriptsize (-8.5\%)} & \multicolumn{2}{c}{} \\
\midrule
\multicolumn{2}{l}{\textbf{Avg. / Improv.}} & & \textbf{5.48e5 {\scriptsize (6.52\%)}} & \textbf{1.57e6 {\scriptsize (27.48\%)}} & \textbf{9.45  {\scriptsize (6.89\%)}} & \textbf{6.19e5 {\scriptsize (-2.89\%)}} & \textbf{1.65e6 {\scriptsize (22.44\%)}} & \textbf{1.09e1 {\scriptsize (-8.45\%)}} & \multicolumn{2}{c}{} \\
\bottomrule
\end{tabular}
}
\begin{tablenotes}
      \small
      \item $^\dagger$ Improvement of \textit{LevelSyn-pwmap} relative to \textit{PigMap-power}.
      \item $^\ddagger$ Improvement of \textit{LevelSyn-pfmap} relative to \textit{PigMap-performance}.
\end{tablenotes}
\end{threeparttable}
\end{table*}

\subsubsection{Mapping Execution and Optimization}

The execution of the physical-informed mapping is implemented as an enhanced flow within the Berkeley ABC engine, extending standard mapping commands such as \textbf{map}, \textbf{pfmap}, and \textbf{pwmap}. During this process, the engine first identifies all potential library cell matches for the subject AIG and evaluates them using the augmented cost function described in Section 3.4.2. By prioritizing matches that align with the predicted spatial trajectory, in particular those that minimize net stretching relative to the fixed PI/PO anchors, the engine ensures a strong correlation between logical structures and physical constraints. This is followed by an iterative refinement phase where LevelSyn performs local gate-level swapping and sizing; a gate is replaced if a functionally equivalent library cell provides a superior fit to the predicted spatial prior, thereby reducing estimated wirelength without violating timing constraints. By embedding physical location directly into the mapping decision tree, LevelSyn generates a netlist that is inherently layout-friendly, significantly mitigating routing congestion and wire-induced delay in the subsequent physical design stages.

%% file: 04_experiment.tex
\section{Experiments} 

In this section, we evaluate the performance of LevelSyn against industry-standard baselines and SOTA physical-aware synthesis methods. We focus on the framework's ability to optimize PPA (Power, Performance, and Area) metrics across a diverse set of benchmarks.

\subsection{Experimental Settings}

\textbf{\quad Benchmark Suite.} We utilize the EPFL Combinational Benchmark Suite \cite{EPFLbenchmark} to assess scalability. The circuit sizes range from hundreds of nodes to over 200k nodes.

\textbf{EDA Flow.}
All designs are processed using Berkeley ABC for logic synthesis and technology mapping. For physical ground truth and spatial anchor generation, we employ DREAMPlace \cite{DreamPlace} to obtain high-quality placement coordinates. The target library is the SkyWater Open Source PDK \cite{skywater_pdk} standard cell library.

\textbf{GNN Training Configuration.}
To ensure the reproducibility of our LA-GNN model, we detail the training environment and hyperparameters as follows: We employ a multi-task loss function $\mathcal{L} =  \mathcal{L}_{MSE}(\hat{x}, \hat{y}) + \mathcal{L}_{BCE}(\hat{O})$ to simultaneously optimize coordinate regression and orientation prediction. $\mathcal{L}_{MSE}$ denotes the Mean Squared Error for the predicted coordinates, while $\mathcal{L}_{BCE}$ represents the Binary Cross-Entropy loss for orientation classification. The model is trained using the AdamW optimizer with an initial learning rate of $1 \times 10^{-3}$ and a weight decay of $1 \times 10^{-4}$. All computational tasks are performed on a Linux workstation equipped with a NVIDIA A100 GPU. We use a hidden dimension of 128 and 6 message-passing layers. The partition window size $w$ is set to 5000 nodes.

\textbf{Baselines.} We employ both the industry-standard baselines and SOTA synthesis methods as baselines, and we address them as follows: 
\begin{itemize}
    \item \textbf{Baseline}: Standard logic-driven mapping in ABC (\textbf{map}).
    \item \textbf{PigMap-power/performance \cite{PigMAP1}}: The SOTA physical-aware mapping that uses analytical wireload models for power and delay optimization.
\end{itemize}

\begin{figure}[t]
    \centering
    \includegraphics[width=0.9\linewidth]{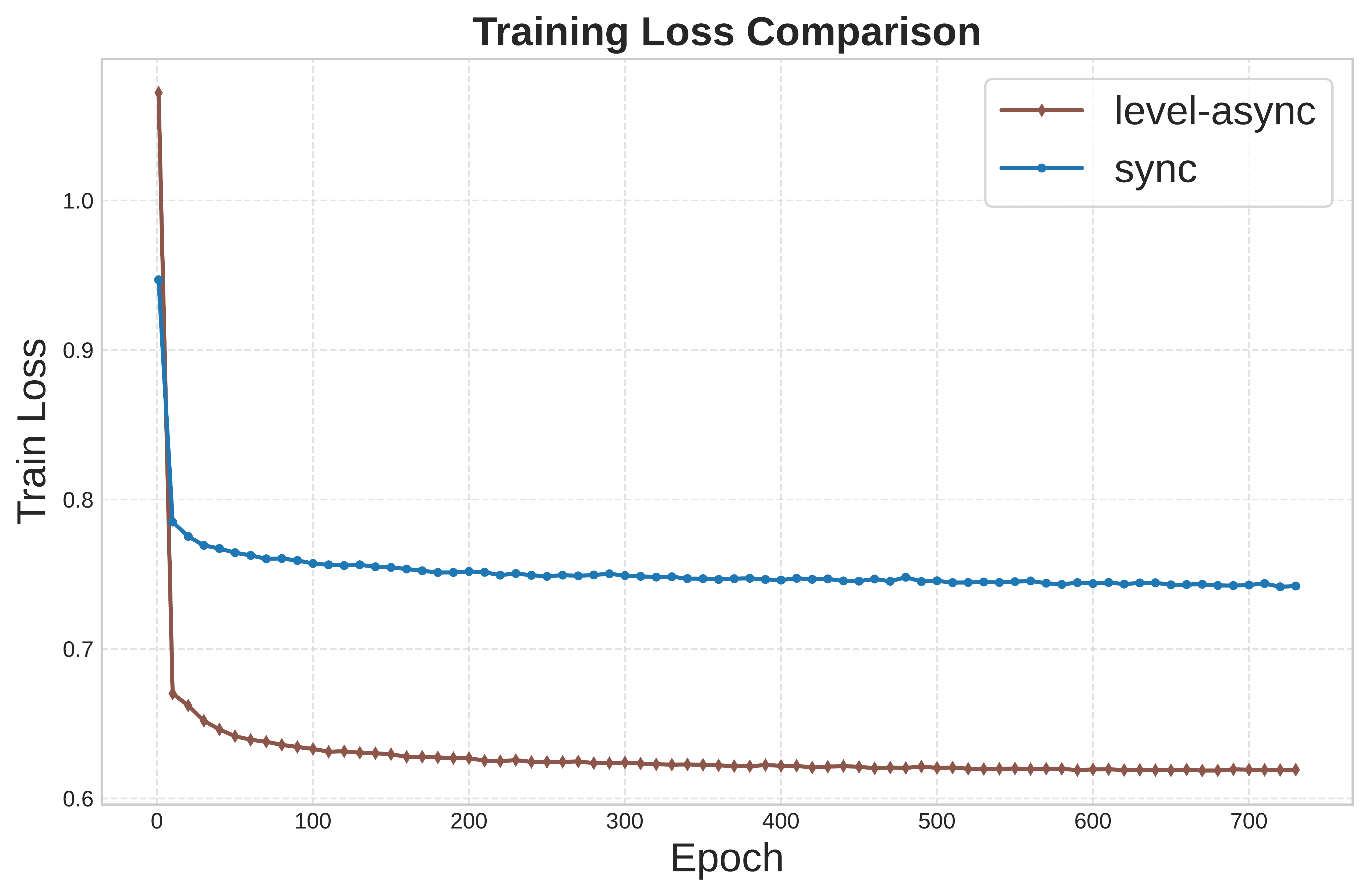}
    \caption{Pre-training loss comparison between LA-GNN and synchronous GNN}
    \label{fig:train_loss_comparison}
    \vspace{-10pt}
\end{figure}

\begin{figure}[t]
    \centering
    \includegraphics[width=0.9\linewidth]{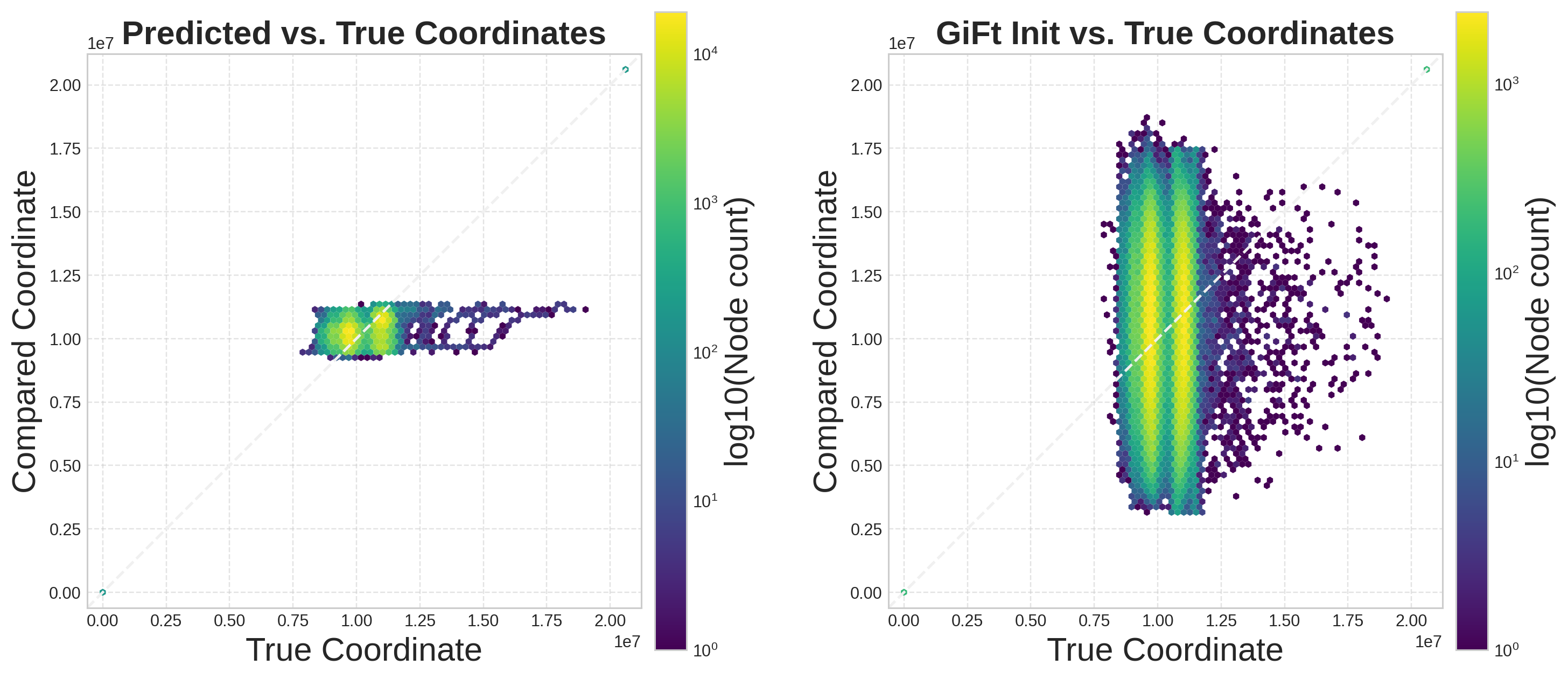}
    \caption{Placement coordinates scatter comparison on hyp. Left: GNN-predicted coordinates. Right: PigMap GiFt\cite{GiFt, PigMAP1} initialization coordinates}
    \label{fig:placement_comparison}
    \vspace{-10pt}
\end{figure}

\subsection{GNN Pre-training and Fine-tuning Results}

To enhance the model's ability to capture generalized structural representations of logic circuits, we adopt a two-stage training strategy consisting of large-scale pre-training followed by task-specific fine-tuning. In the pre-training phase, we construct a diverse dataset comprising 500 to 10,000 randomly generated AIGs with varying topological complexities, which allows the LA-GNN to learn intrinsic geometric and logical features across a broad synthetic design space. Subsequently, the pre-trained weights are used as an initialization for the model, which is then fine-tuned on specific circuits from established benchmarks to adapt the learned representations to realistic logic synthesis and physical design constraints. This transfer learning approach not only accelerates the convergence process but also significantly improves the model's predictive accuracy and robustness when encountering the structural variations inherent in different circuit categories.

We evaluated the convergence behavior of our proposed LA-GNN (level-async) approach against a standard synchronous (sync) updating scheme (5-layer GCN) over 700 epochs. The training results are shown in Figure \ref{fig:train_loss_comparison}. Both methods exhibit a rapid reduction in training loss during the initial phase. However, the level-async strategy demonstrates a markedly superior optimization trajectory. While the standard sync method settles at a higher training loss of ~0.75, the level-async exhibits a sharper initial descent and converges to a significantly lower loss of 0.61.

We visualize the coordinate regression performance  in Figure \ref{fig:placement_comparison} on the largest benchmark circuit, hyp, which comprises 214,591 nodes. The scatter plots compare the GNN-predicted coordinates (left) and the PigMap GiFt \cite{GiFt, PigMAP1} initialization coordinates (right) against the true ground-truth placements. While the GiFt initialization results in a highly chaotic and scattered distribution, our LA-GNN effectively suppresses this noise and concentrate the predicted coordinates into a refined and high-density manifold. This focused prediction suggests that the LA-GNN has successfully learned the intrinsic logical connectivity and global center of gravity of the circuit, rather than falling into the high-variance irregularities seen in GiFt. By providing such a stable and cohesive initial layout, our model offers a much more reliable and high-quality starting point for downstream placement optimization, which is crucial for achieving better wirelength and timing convergence in ultra-large-scale designs.

\begin{table}[t]
\centering
\caption{Post-Routing Results on Bar}
\label{tab:power_comparison}
\resizebox{\columnwidth}{!}{
\begin{tabular}{lccc}
\toprule
 \textbf{Strategies} & \textbf{Wire Length (um)} & \textbf{Via Counts}  & \textbf{DRC Violations}  \\ 
\midrule
\textbf{\textit{PigMap-power}} &320,689  & 33,963 & 966  \\
\textbf{\textit{PigMap-performance}} &301,312 & 28,025 & 623  \\
\textbf{\textit{LevelSyn-pwmap}} &254,439 & 24,330 & 4 \\
\textbf{\textit{LevelSyn-pfmap}} &243,084 & 22,419 & 17 \\
\midrule
\textbf{$\mathcal{}{Imp.}^\dagger$}  & 20.66\% & 28.36\% & 99.59\% \\
\textbf{$\mathcal{}{Imp.}^\ddagger$}  & 19.32\% & 20.00\% & 97.27\% \\
\bottomrule
\end{tabular}
}
\begin{tablenotes}
      \small
      \item $^\dagger$ Improvement of \textit{LevelSyn-pwmap} relative to \textit{PigMap-power}.
      \item $^\ddagger$ Improvement of \textit{LevelSyn-pfmap} relative to \textit{PigMap-performance}.
\end{tablenotes}
\end{table}

\subsection{PPA Results}
To evaluate the effectiveness of the proposed framework under different optimization goals, we implement two variants of LevelSyn: \textit{LevelSyn-pwmap} and \textit{LevelSyn-pfmap}. The former employs \textbf{pwmap} to minimize total wirelength and switching power, while the latter utilizes \textbf{pfmap} to focus on critical path delay. Table \ref{tab:main_result} summarizes the PPA results. LevelSyn consistently outperforms both the logic-only baseline and PigMap across the majority of test cases.

Experimental results demonstrate that LevelSyn consistently outperforms PigMap across different optimization goals: specifically, \textit{LevelSyn-pwmap} achieves an average 6.52\% area reduction and 6.89\% switching power improvement compared to \textit{PigMap-power}, with notable gains in large designs like \textit{hyp}. These results demonstrate the accuracy of wirelength estimation provided by LA-GNN. Meanwhile, the delay-oriented \textit{LevelSyn-pfmap} reduces delay by 22.44\% by effectively shortening critical paths through Manhattan distance optimization; remarkably, \textit{LevelSyn-pwmap} yields the best overall delay reduction of 27.48\%, suggesting that our global spatial anchors and NLI-based positioning offer such high-fidelity spatial guidance that they significantly enhance global structural alignment and timing outcomes beyond traditional logic-depth-based heuristics.

To further evaluate the practical efficacy of our approach within a complete physical design flow, we conduct a comprehensive case study on the Bar circuit, and all strategies are subjected to exactly 20 iterations of routing optimization. The choice of Bar as our primary subject is deliberate: it serves as the central case study in SOTA baseline PigMap.

As summarized in Table~\ref{tab:power_comparison}, our LevelSyn framework demonstrates a decisive advantage over the baseline across all critical metrics. Specifically, when comparing our power-optimized variant (\textit{LevelSyn-pwmap}) to \textit{PigMap-power}, we achieve a 20.66\% reduction in total wire length and a 28.36\% decrease in via counts. Most notably, our method achieves a near-total elimination of Design Rule Check (DRC) violations, plummeting from 966 to just 4, which is a staggering 99.59\% improvement. Similarly, our performance-oriented variant (LevelSyn-pfmap) outperforms its counterpart with a 19.32\% improvement in wire length and a 97.27\% reduction in DRC violations. These results confirm that with the same 20-round optimization budget, the structural stability provided by our LA-GNN allows the router to converge on a significantly more robust, and manufacturable layout.

\vspace{-10pt}
\begin{table*}[t]
\centering
\caption{Runtime Comparison}
\label{tab:runtime_analysis}
\resizebox{\linewidth}{!}{
\begin{tabular}{lcccccc}
\toprule
\textbf{Circuit Name}s & \textbf{Nodes} & \textbf{Baseline} ($s$) & \textbf{\textit{PigMAP-power}} ($s$) & \textbf{\textit{PigMAP-performance}} ($s$) & \textbf{\textit{LevelSyn-pwmap}} ($s$) & \textbf{\textit{LevelSyn-pfmap}} ($s$) \\ 
\midrule
ctrl       & 181    & 0.25176  & 4.26475 (0.26475+4)      & 4.26010 (0.26010+4)       & 0.34770 (0.29462+0.05308)       & 0.30497 (0.25189+0.05308)      \\
int2float  & 271    & 0.24799  & 3.27121 (0.27121+3)      & 3.28982 (0.28982+3)     & 0.35021 (0.27180+0.07841)      & 0.35164 (0.27323+0.07841)      \\
dec        & 312    & 0.26014  & 3.28175 (0.28175+3)      & 3.28115 (0.28115+3)     & 0.3237 (0.31036+0.01334)       & 0.33829 (0.32495+0.01334)      \\
router     & 317    & 0.25183  & 5.27981 (0.27981+5)      & 5.28016 (0.28016+5)     & 0.43717 (0.28035+0.15682)      & 0.43670 (0.27988+0.15682)       \\
cavlc      & 703    & 0.28075  & 3.32366 (0.32366+3)      & 3.32096 (0.32096+3)     & 0.38326 (0.30978+0.07348)      & 0.40011 (0.32663+0.07348)      \\
priority   & 1106   & 0.26983  & 5.34651 (0.34651+5)      & 5.35716 (0.35716+5)     & 1.33215 (0.35595+0.97620)      & 1.33603 (0.35983+0.97620)      \\
adder      & 1276   & 0.26596  & 5.36133 (0.36133+5)      & 5.36947 (0.36947+5)     & 1.46657 (0.36737+1.09920)      & 1.46161 (0.36241+1.09920)      \\
i2c        & 1489   & 0.28782  & 4.41634 (0.41634+4)      & 4.41147 (0.41147+4)     & 0.51008 (0.42255+0.08753)      & 0.52417 (0.43664+0.08753)      \\
max        & 3377   & 0.31464  & 6.08305 (2.08305+4)      & 5.65394 (1.65394+4)     & 2.81013 (1.68296+1.12717)      & 2.83736 (1.71019+1.12717)      \\
bar        & 3471   & 0.29879  & 4.06734 (1.06734+3)      & 4.04371 (1.04371+3)   & 1.27092 (1.05667+0.21425)      & 1.39967 (1.18542+0.21425)      \\
sin        & 5440   & 0.53017  & 6.58373 (2.58373+4)      & 6.66237 (2.66237+4)     & 3.43735 (2.61093+0.82642)      & 3.60141 (2.77499+0.82642)      \\
arbiter    & 12095  & 0.54331  & 22.62736 (16.62736+6)    & 20.33924 (14.33924+6)   & 16.93909 (16.41279+0.52630)    & 13.79946 (13.27316+0.52630)    \\
voter      & 14759  & 0.80098  & 20.50558 (15.50558+5)    & 20.87620 (15.87620+5)     & 15.82894 (15.43108+0.39786)    & 15.88984 (15.49198+0.39786)    \\
square     & 18548  & 0.95917  & 23.97164 (18.97164+5)    & 24.19507 (19.19507+5)   & 20.14333 (18.48149+1.66184)    & 20.22944 (18.56760+1.66184)     \\
sqrt       & 24746  & 1.11424  & 32.76447 (26.76447+6)    & 33.58574 (27.58574+6)   & 58.05863 (27.36181+30.69682)   & 57.99451 (27.29769+30.69682)   \\
multiplier & 27190  & 1.31417  & 47.66312 (40.66312+7)    & 47.80954 (40.80954+7)   & 43.68709 (41.51560+2.17149)     & 42.88535 (40.71386+2.17149)    \\
log2       & 32092  & 1.77527  & 60.93639 (52.93639+8)    & 61.99767 (53.99767+8)   & 57.04942 (54.20919+2.84023)    & 57.19582 (54.35559+2.84023)    \\
mem\_ctrl  & 48040  & 1.33803  & 167.97727 (158.97727+9)  & 161.89096 (152.89096+9) & 154.33898 (153.26823+1.07075)  & 152.5755 (151.50475+1.07075)   \\
div        & 57375  & 2.72399  & 290.26830 (281.26830+9)    & 270.12475 (261.12475+9) & 318.20752 (269.53765+48.66987) & 289.11868 (240.44881+48.66987) \\
hyp        & 214591 & 11.33212 & 968.93256 (943.93256+25) & 970.12293 (945.12293+25)& 1,647.60024 (904.50277+743.09747)& 1,726.23648 (983.13901+743.09747)\\ 
\bottomrule
\end{tabular}
}
\end{table*}
\subsection{Runtime Analysis}
The runtime analysis is presented in Table \ref{tab:runtime_analysis}. In the PigMAP and LevelSyn columns, the parenthetical values represent the breakdown of runtime as (Mapping Time + Pre-processing/Training Time). Specifically, for PigMAP, this includes the GiFt\cite{GiFt} initialization time, whereas for LevelSyn, it accounts for the specific model fine-tuning time. It demonstrates that the LevelSyn approach maintains high efficiency while achieving superior mapping results. Although our methodology includes a comprehensive training strategy, the overall execution time remains highly competitive with established methods like PigMAP. 

The empirical data indicates that our specific training approach does not lead to the significant runtime extensions, which is often associated with deep learning applications in EDA. In several cases such as the multiplier and mem\_ctrl circuits, the total runtime of LevelSyn is even comparable to or slightly lower than that of PigMAP. This proves that the fine-tuning phase is remarkably efficient across various circuit scales. Furthermore, while the mapping duration for certain large scale circuits like sqrt and hyp shows a relative increase compared to baseline methods, this latency is negligible when evaluated within the context of the entire electronic design automation flow. In actual industrial practice, the subsequent physical design stages including placement and routing typically require several hours or even days to complete. Consequently, the marginal extra time spent during the logic mapping stage is a minor trade-off for the substantial improvements in power and performance metrics provided by LevelSyn. By producing a more optimized netlist, our approach likely eases the burden on downstream back-end tools and facilitates faster timing closure.
\subsection{Ablation Study}

To evaluate the memory efficiency of our proposed LA-GNN architecture, we conduct an ablation study comparing its peak GPU memory consumption against a 5-layer standard GCN baseline. As summarized in Table~\ref{tab:ablation_study}, the results reveal a significant crossover in memory performance as the circuit scale increases.

Inherently, the LA-GNN architecture requires more memory than a standard GCN due to the additional overhead of maintaining level-specific states and managing asynchronous message-passing buffers. This is evident in several smaller benchmarks where LA-GNN’s allocated memory slightly exceeds that of the GCN. However, this architectural overhead is effectively mitigated by our LASP Strategy as we move toward large-scale designs.

The standard GCN suffers from poor scalability because it requires simultaneous storage of the entire graph’s adjacency structure and gradient states, leading to a memory explosion on industrial-sized circuits. In contrast, our partitioning strategy decomposes massive AIGs into optimized sub-graphs, allowing the model to process level-dependent information without overloading the GPU's VRAM. For the largest benchmark, hyp, this synergy allows LA-GNN to use only 4,709 MB of allocated memory compared to the GCN’s 8,956 MB, which is a reduction of approximately 47.4\%. These results demonstrate that while LAGNN is more complex by design, the integration of our sub-graph partitioning makes it a far more scalable and practical solution for ultra-large-scale circuit placement, where memory constraints are most critical.
\begin{table}[htbp]
\vspace{-10pt}
\centering
\caption{Peak GPU Memory Comparison}
\label{tab:ablation_study}
\resizebox{\columnwidth}{!}{
\begin{tabular}{lcrrrr}
\toprule
\multirow{2}{*}{Circuit Name} & \multirow{2}{*}{Nodes}& \multicolumn{2}{c}{\textbf{LA-GNN}} & \multicolumn{2}{c}{\textbf{GCN}} \\
\cmidrule(lr){3-4}\cmidrule(lr){5-6}
& &allocated ($MB$) & reserved ($MB$)
&  allocated ($MB$)&  reserved ($MB$) \\
\midrule
ctrl       & 181 &28.209 & 54.000   & 25.272   & 50.000    \\
int2float  & 271 &35.948 & 62.000   & 28.490   & 54.000    \\
router     & 312 &51.848 & 76.000   & 31.605   & 58.000    \\
dec        & 317 &30.314 & 54.000   & 30.594   & 56.000    \\
cavlc      & 703 &68.501 & 88.000   & 47.588   & 72.000    \\
priority   & 1106 &308.150& 332.000  & 68.157   & 90.000    \\
i2c        & 1276& 125.399& 132.000  & 75.263   & 90.000    \\
adder      & 1489 &391.053& 406.000  & 82.287   & 110.000   \\
bar        & 3377& 89.509 & 110.000  & 153.047  & 178.000   \\
max        & 3471&291.921& 336.000  & 173.920  & 220.000   \\
sin        & 5440& 241.344& 250.000  & 218.373  & 272.000   \\
arbiter    & 12095&299.112& 604.000  & 547.051  & 616.000   \\
voter      & 14759&340.547& 678.000  & 650.676  & 716.000   \\
square     & 18548&431.628& 890.000  & 815.992  & 880.000   \\
sqrt       & 24746&1315.981& 2080.000& 923.094  & 1024.000  \\
multiplier & 27190&570.486& 1080.000 & 1091.725 & 1242.000  \\
log2       & 32092&621.266& 1428.000 & 1200.106 & 1384.000  \\
mem\_ctrl  & 48040&992.337& 1792.000 & 1911.500 & 2156.000  \\
div        & 54375&1259.288& 2828.000& 2258.522 & 2490.000  \\
hyp        & 214591&4709.544& 9604.000& 8956.849 & 10668.000 \\
\bottomrule
\end{tabular}
}
\end{table}

%% file: 05_conclusion.tex
\section{Conclusion and Future Work}\label{Sec:Conclusion}
In conclusion, this paper presents PhySyn, a framework that successfully bridges the gap between logic synthesis and physical implementation by embedding deep learning-based spatial predictions into the synthesis flow. Through the use of LA-GNN and level-aligned partitioning, PhySyn achieves high-fidelity spatial estimation and scalability for large-scale designs. Specifically, the near-total elimination of DRC violations underscores the potential of physical-aware synthesis in achieving correct-by-construction design flows. Moving forward, we plan to extend this framework to support sequential circuits by incorporating flip-flop positioning and to explore multi-objective optimization targets such as congestion and thermal distribution.

%% file: reference.bib
@String{Computing = "Computing" }

@String{Computer = "{IEEE} Computer" }

@String{Academic = "Academic Press" }

@String{Springer = "Springer-Verlag" }

@Inproceedings{li2022deepgate,
  title="Deepgate: Learning neural representations of logic gates",
  author="Li, Min and Khan, Sadaf and Shi, Zhengyuan and Wang, Naixing and Yu, Huang and Xu, Qiang",
  booktitle="Proceedings of the 59th ACM/IEEE Design Automation Conference",
  pages="667--672",
  year="2022"
}

@inproceedings{brayton2010abc,
  title={ABC: An academic industrial-strength verification tool},
  author={Brayton, Robert and Mishchenko, Alan},
  booktitle={Computer Aided Verification},
  pages={24--40},
  year={2010},
  organization={Springer}
}

@inproceedings{wolf2013yosys,
  title={Yosys-a free Verilog synthesis suite},
  author={Wolf, Clifford and Glaser, Johann and Kepler, Johannes},
  booktitle={Proceedings of the 21st Austrian Workshop on Microelectronics (Austrochip)},
  pages={97},
  year={2013}
}

@inproceedings{shi2023deepgate2,
  title={DeepGate2: Functionality-Aware Circuit Representation Learning},
  author={Shi, Zhengyuan and Pan, Hongyang and Khan, Sadaf and Li, Min and Liu, Yi and Huang, Junhua and Zhen, Hui-Ling and Yuan, Mingxuan and Chu, Zhufei and Xu, Qiang},
  booktitle={Proceedings of the 2023 IEEE/ACM international conference on Computer-aided design},
  year={2023}
}

@inproceedings{liu24polargate,
    author = {Jiawei Liu and Jianwang Zhai and Mingyu Zhao and Zhe Lin and Bei Yu and Chuan Shi},
    title = {PolarGate: Breaking the Functionality Representation Bottleneck of And-Inverter Graph Neural Network},
    booktitle = {Proceedings of the 43rd IEEE/ACM International Conference on Computer-Aided Design},
    year = {2024}
}

@article{mishchenko2007abc,
  title={ABC: A system for sequential synthesis and verification},
  author={Mishchenko, Alan and others},
  journal={URL http://www. eecs. berkeley. edu/alanmi/abc},
  volume={17},
  year={2007}
}

@article{chen2024large,
  title={Large circuit models: opportunities and challenges},
  author={Chen, Lei and Chen, Yiqi and Chu, Zhufei and Fang, Wenji and Ho, Tsung-Yi and Huang, Ru and Huang, Yu and Khan, Sadaf and Li, Min and Li, Xingquan and others},
  journal={Science China Information Sciences},
  volume={67},
  number={10},
  pages={200402},
  year={2024},
  publisher={Springer}
}

@article{wu2025shift,
  title={Shift-Left Techniques in Electronic Design Automation: A Survey},
  author={Wu, Xinyue and Li, Zixuan and Hu, Fan and Lin, Ting and Zhao, Xiaotian and Wang, Runxi and Guo, Xinfei},
  journal={arXiv preprint arXiv:2509.14551},
  year={2025}
}

@inproceedings{pedram1991layout,
  title={Layout driven technology mapping},
  author={Pedram, Massoud and Bhat, Narasimha},
  booktitle={Proceedings of the 28th ACM/IEEE Design Automation Conference},
  pages={99--105},
  year={1991}
}

@inproceedings{lu1998combining,
  title={Combining technology mapping with post-placement resynthesis for performance optimization},
  author={Lu, Aiguo and Eisenmann, Hans and Stenz, Guenter and Johannes, Frank M},
  booktitle={Proceedings International Conference on Computer Design. VLSI in Computers and Processors (Cat. No. 98CB36273)},
  pages={616--621},
  year={1998},
  organization={IEEE}
}

@inproceedings{xie2021net2,
  title={Net2: A graph attention network method customized for pre-placement net length estimation},
  author={Xie, Zhiyao and Liang, Rongjian and Xu, Xiaoqing and Hu, Jiang and Duan, Yixiao and Chen, Yiran},
  booktitle={Proceedings of the 26th Asia and South Pacific Design Automation Conference},
  pages={671--677},
  year={2021}
}

@inproceedings{wu2025differentiable,
  title={Differentiable Graph Neural Networks for Wirelength Estimation},
  author={Wu, Zhengfeng and Shrestha, Pratik and Phatharodom, Saran and Savidis, Ioannis},
  booktitle={2025 IEEE International Symposium on Circuits and Systems (ISCAS)},
  pages={1--5},
  year={2025},
  organization={IEEE}
}

@inproceedings{he2025cpa,
  title={CPA-Remap: Critical-Path-Based Physically Aware Remapping Framework for Timing Optimization},
  author={He, Mingxiao and Huang, Pengcheng and Zhao, Zhenyu and Bian, Peiyun},
  booktitle={2025 IEEE 43rd International Conference on Computer Design (ICCD)},
  pages={393--400},
  year={2025},
  organization={IEEE}
}

@inproceedings{deng2024less,
  title={Less is more: Hop-wise graph attention for scalable and generalizable learning on circuits},
  author={Deng, Chenhui and Yue, Zichao and Yu, Cunxi and Sarar, Gokce and Carey, Ryan and Jain, Rajeev and Zhang, Zhiru},
  booktitle={Proceedings of the 61st ACM/IEEE Design Automation Conference},
  pages={1--6},
  year={2024}
}

@article{fang2025nettag,
  title={NetTAG: A Multimodal RTL-and-Layout-Aligned Netlist Foundation Model via Text-Attributed Graph},
  author={Fang, Wenji and Li, Wenkai and Liu, Shang and Lu, Yao and Zhang, Hongce and Xie, Zhiyao},
  journal={arXiv preprint arXiv:2504.09260},
  year={2025}
}

@inproceedings{shi22deeptpi,
    author = {Zhengyuan Shi and Min Li and Sadaf Khan and Liuzheng Wang and Naixing Wang and Yu Huang},
    title = {DeepTPI: Test Point Insertion with Deep Reinforcement Learning},
    booktitle = {2022 IEEE International Test Conference (ITC)},
    page = {194–203},
    year = {2022}
}

@inproceedings{Limin23sat,
    author = {Min Li and Zhengyuan Shi and Qiuxia Lai and Sadaf Khan and Shaowei Cai and Qiang Xu.},
    title = {On eda-driven learning for sat solving},
    booktitle = { Proceedings of the 60th Annual ACM/IEEE Design Automation Conference},
    page ={1--6},
    year = {2023}
}

@inproceedings{fang2025circuitfusion,
  title        = {CircuitFusion: Multimodal Circuit Representation Learning for Agile Chip Design},
  author       = {Wenji Fang and Shang Liu and Jing Wang and Zhiyao Xie},
  booktitle    = {International Conference on Learning Representations (ICLR)},
  year         = {2025}
}

@inproceedings{PigMAP1,
author = {Pan, Hongyang and Lan, Cunqing and Liu, Yiting and Wang, Zhiang and Shang, Li and Zeng, Xuan and Yang, Fan and Zhu, Keren},
title = {Physically Aware Synthesis Revisited: Guiding Technology Mapping with Primitive Logic Gate Placement},
year = {2025},
booktitle = {Proceedings of the 43rd IEEE/ACM International Conference on Computer-Aided Design},
articleno = {171}
}

@ARTICLE{PigMAP2,
  author={Lan, Cunqing and Pan, Hongyang and Wang, Zhiang and Zeng, Xuan and Yang, Fan and Zhu, Keren},
  journal={IEEE Transactions on Computer-Aided Design of Integrated Circuits and Systems}, 
  title={PigMap2: A Physical Information Guided Technology Mapping Framework}, 
  year={2025},
  }

@inproceedings{GiFt,
author = {Liu, Yiting and Zhou, Hai and Wang, Jia and Yang, Fan and Zeng, Xuan and Shang, Li},
title = {The Power of Graph Signal Processing for Chip Placement Acceleration},
year = {2025},
publisher = {Association for Computing Machinery},
booktitle = {Proceedings of the 43rd IEEE/ACM International Conference on Computer-Aided Design},
series = {ICCAD '24}
}

@ARTICLE{future-of-wires,
  author={Ho, R. and Mai, K.W. and Horowitz, M.A.},
  journal={Proceedings of the IEEE}, 
  title={The future of wires}, 
  year={2001},
  volume={89},
  number={4},
  pages={490-504},
}

@INPROCEEDINGS{Address_problem,
  author={Gosti, W. and Khatri, S.R. and Sangiovanni-Vincentelli, A.L.},
  booktitle={IEEE/ACM International Conference on Computer Aided Design. ICCAD 2001. IEEE/ACM Digest of Technical Papers (Cat. No.01CH37281)}, 
  title={Addressing the timing closure problem by integrating logic optimization and placement}, 
  year={2001}
  }

@inproceedings{EPFLbenchmark,
  title={The EPFL Combinational Benchmark Suite},
  author={Luca Gaetano Amar{\`u} and Pierre-Emmanuel Gaillardon and Giovanni De Micheli},
  year={2015},
  url={https://api.semanticscholar.org/CorpusID:13971503}
}

@ARTICLE{delay_opt_fpga,
  author={Chau-Shen Chen and Yu-Wen Tsay and TingTing Hwang and Wu, A.C.H. and Youn-Long Lin},
  journal={IEEE Transactions on Computer-Aided Design of Integrated Circuits and Systems}, 
  title={Combining technology mapping and placement for delay-minimization in FPGA designs}, 
  year={1995}
  }

@INPROCEEDINGS{placement_fpga,
  author={Tong GaO and Kuang-Chien Chen and Cong, J. and Yuzheng Ding and Liu, C.},
  booktitle={Sixth Annual IEEE International ASIC Conference and Exhibit}, 
  title={Placement and placement driven technology mapping for FPGA synthesis}, 
  year={1993}}

@inproceedings{dsm_design_flow,
author = {Salek, Amir H. and Lou, Jinan and Pedram, Massoud},
title = {A DSM design flow: putting floorplanning, technology-mapping, and gate-placement together},
year = {1998},
booktitle = {Proceedings of the 35th Annual Design Automation Conference},
series = {DAC '98}
}

@Inbook{Reis2018,
author="Reis, Andr{\'e} In{\'a}cio
and Matos, Jody M. A.",
editor="Reis, Andr{\'e} In{\'a}cio
and Drechsler, Rolf",
title="Physical Awareness Starting at Technology-Independent Logic Synthesis",
bookTitle="Advanced Logic Synthesis"
}

@inproceedings{Cluster-GNN,
author = {Chiang, Wei-Lin and Liu, Xuanqing and Si, Si and Li, Yang and Bengio, Samy and Hsieh, Cho-Jui},
title = {Cluster-GCN: An Efficient Algorithm for Training Deep and Large Graph Convolutional Networks},
year = {2019},
isbn = {9781450362016},
booktitle = {Proceedings of the 25th ACM SIGKDD International Conference on Knowledge Discovery \& Data Mining},
pages = {257–266},
numpages = {10}
}

@article{graph-partition,
author = {Karypis, George and Kumar, Vipin},
title = {A Fast and High Quality Multilevel Scheme for Partitioning Irregular Graphs},
year = {1998},
month = dec,
pages = {359–392},
numpages = {34}
}

@ARTICLE{DreamPlace,
  author={Lin, Yibo and Jiang, Zixuan and Gu, Jiaqi and Li, Wuxi and Dhar, Shounak and Ren, Haoxing and Khailany, Brucek and Pan, David Z.},
  journal={IEEE Transactions on Computer-Aided Design of Integrated Circuits and Systems}, 
  title={DREAMPlace: Deep Learning Toolkit-Enabled GPU Acceleration for Modern VLSI Placement}, 
  year={2021}}

@misc{skywater_pdk,
  author = {{SkyWater Technology and Google}},
  title = {{SkyWater Open Source PDK}},
  year = {2020},
  url = {https://github.com/google/skywater-pdk},
  note = {Accessed: 2026-04-05}
}

@INPROCEEDINGS{togawa1994maple,
  author={Togawa, N. and Sato, M. and Ohtsuki, T.},
  booktitle={IEEE/ACM International Conference on Computer-Aided Design}, 
  title={Maple: A Simultaneous Technology Mapping, Placement, And Global Routing Algorithm For Field-programmable Gate Arrays}, 
  year={1994},
  pages={156-163},
  doi={10.1109/ICCAD.1994.629759}}

@ARTICLE{togawa1998maple,
  author={Togawa, N. and Yanagisawa, M. and Ohtsuki, T.},
  journal={IEEE Transactions on Computer-Aided Design of Integrated Circuits and Systems}, 
  title={Maple-opt: a performance-oriented simultaneous technology mapping, placement, and global routing algorithm for FPGAs}, 
  year={1998},
  pages={803-818},
  doi={10.1109/43.720317}}

@misc{openroad_repo,
  author       = {{The OpenROAD Project}},
  title        = {{OpenROAD}: An open-source industrial design flow from {RTL} to {GDS}},
  year         = {2024},
  publisher    = {GitHub},
  howpublished = {\url{https://github.com/The-OpenROAD-Project/OpenROAD}}
}

@article{chen2024dawn,
  title={The dawn of ai-native eda: Opportunities and challenges of large circuit models},
  author={Chen, Lei and Chen, Yiqi and Chu, Zhufei and Fang, Wenji and Ho, Tsung-Yi and Huang, Ru and Huang, Yu and Khan, Sadaf and Li, Min and Li, Xingquan and others},
  journal={arXiv preprint arXiv:2403.07257},
  year={2024}
}

@article{huang2021machine,
  title={Machine learning for electronic design automation: A survey},
  author={Huang, Guyue and Hu, Jingbo and He, Yifan and Liu, Jialong and Ma, Mingyuan and Shen, Zhaoyang and Wu, Juejian and Xu, Yuanfan and Zhang, Hengrui and Zhong, Kai and Ning, Xuefei and Ma, Yuzhe and Yang, Haoyu and Yu, Bei and Yang, Huazhong and Wang, Yu},
  journal = {ACM Trans. Des. Autom. Electron. Syst.},
  year={2021}
}
